\documentclass[12pt]{article}
\usepackage{times}
\usepackage{geometry}
\usepackage[utf8]{inputenc}
\usepackage{enumitem,amssymb}
\usepackage{ragged2e}
\usepackage[svgnames]{xcolor}

\usepackage[hidelinks]{hyperref}
\hypersetup{
     colorlinks = true,
     linkcolor = teal,
     anchorcolor = blue,
     citecolor = teal,
     filecolor = blue,
     urlcolor = blue
     }

\usepackage[numbers]{natbib}
\usepackage{aastex_hack}
\usepackage{graphics,graphicx}

\newlist{thematic}{itemize}{8}
\setlist[thematic]{label=$\square$}
\usepackage{pifont}
\newcommand{\cmark}{\ding{51}}%
\newcommand{\done}{\rlap{$\square$}{\raisebox{2pt}{\large\hspace{1pt}\cmark}}%
\hspace{-2.5pt}}

\newcommand{\nicer}{\textit{NICER}}

\begin{document}
\raggedright
\huge
Astro2020 Science White Paper \linebreak

Determining the Equation of State of Cold, Dense Matter with X-ray Observations of Neutron Stars \linebreak
\normalsize

\noindent \textbf{Thematic Areas:} \hspace*{60pt} $\square$ Planetary Systems \hspace*{10pt} $\square$ Star and Planet Formation \hspace*{20pt}\linebreak
\done\ Formation and Evolution of Compact Objects \hspace*{31pt} \done\ Cosmology and Fundamental Physics \linebreak
  $\square$  Stars and Stellar Evolution \hspace*{1pt} $\square$ Resolved Stellar Populations and their Environments \hspace*{40pt} \linebreak
  $\square$    Galaxy Evolution   \hspace*{45pt} $\square$             Multi-Messenger Astronomy and Astrophysics \hspace*{65pt} \linebreak
  
\textbf{Principal Author:}

Name:	Slavko Bogdanov
 \linebreak						
Institution:  Columbia University
 \linebreak
Email: \href{mailto:slavko@astro.columbia.edu}{slavko@astro.columbia.edu} 
 \linebreak
 
\textbf{Co-authors:}
  \linebreak
  Anna L.~Watts, \textit{University of Amsterdam}\\
  Deepto Chakrabarty, \textit{Massachussetts Institute of Technology}\\
  Zaven Arzoumanian, \textit{NASA Goddard Space Flight Center}\\
  Sebastien Guillot, \textit{IRAP and Universit\'{e} de Toulouse}\\
  Keith C.~Gendreau, \textit{NASA Goddard Space Flight Center}\\
  Frederick K.~Lamb, \textit{University of Illinois at Urbana-Champaign}\\
  Thomas Maccarone, \textit{Texas Tech University}\\
  M.~Coleman Miller, \textit{University of Maryland}\\
  Feryal~\"Ozel, \textit{University of Arizona}\\
  Paul S.~Ray, \textit{Naval Research Laboratory}\\
  Colleen A.~Wilson-Hodge, \textit{NASA Marshall Space Flight Center}
  \linebreak

\justify
\textbf{Abstract:}
\linebreak
The unknown state of matter at ultra-high density, large proton/neutron number asymmetry, and low temperature is a major long-standing problem in modern physics.  Neutron stars provide the only known setting in the Universe where matter in this regime can stably exist. Valuable information about the interior structure of neutron stars can be extracted via sensitive observations of their exteriors. There are several complementary techniques that require different combinations of high time resolution, superb spectral resolution, and high spatial resolution. In the upcoming decade and beyond, measurements of the masses and radii of an ensemble of neutron stars using these techniques, based on data from multiple proposed next-generation X-ray telescopes, can produce definitive empirical constraints on the allowed dense matter equation of state.

\pagebreak

\justify

\vspace{-0.2cm}
\section{Introduction}
\vspace{-0.2cm}
The lack of knowledge about the physical properties of cold, stable matter at densities that exceed the nuclear saturation density ($\rho_s=2.8 \times 10^{14}$\,g\,cm$^{-3}$) remains one of the principal outstanding problems in nuclear physics, owing to a number of challenges both in the experimental and theoretical realms \cite[see, e.g.,][for a review]{Watts16}. A number of plausible theoretical predictions for the state of matter in this regime exist, which range from normal nucleonic matter, to particle exotica such as hyperons, deconfined quarks, color superconducting phases, and Bose-Einstein condensates. Matter in this extreme regime of  ultra-high density, large proton/neutron number asymmetry, and low temperature can only exist stably in the cores of neutron stars (NSs), which makes them of tremendous value for nuclear physics as they serve as important natural laboratories for studying the physics of the strong interaction and the state of supranuclear matter. Determining the dense matter equation of state (EoS) has far-reaching implications for astrophysics as well.  The detailed physics and the accompanying electromagnetic, neutrino, and gravitational wave signals of some of the most energetic phenomena in the Universe, such as black-hole/NS and double NS mergers \citep{Metzger_2010,Kumar15} as well as core-collapse supernovae, are highly sensitive to the interior structure of NSs \cite{Janka_2012}. 

\begin{figure}[tb]
\centering
\includegraphics[width=0.94\textwidth]{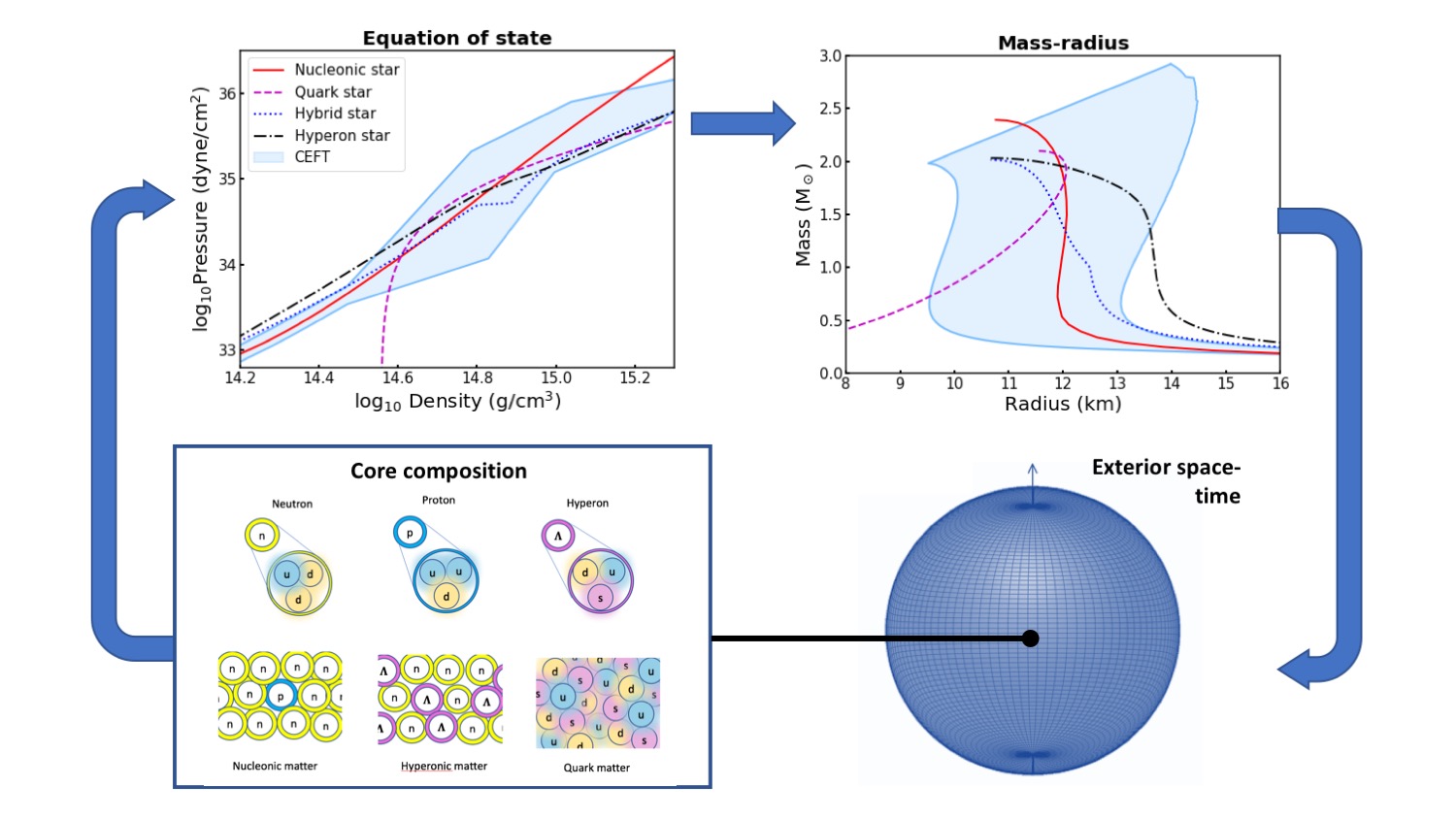}
\caption{\small{The particle content and their interactions in the high density -- low temperature setting at the cores of NSs is highly uncertain.  Our lack of knowledge about these microphysical aspects (lower left:  uds = up down strange quarks) is encapsulated in the EoS (top left).  A sampling of plausible theoretical EoS models is shown that includes a nucleonic star (red) \cite{Akmal97}, a quark star (magenta) \cite{Li16}, a hybrid star consisting of a nucleonic outer core and quark matter inner core (blue) \cite{Zdunik13}, and a hyperon star with nucleonic outer core and hyperonic matter inner core (black) \cite{Bednarek_etal_2012}. The light blue region represents the approximate range spanned by the set of currently viable models \cite{hebeler13}.  The different EoS govern the global properties of the star such as $M$, $R$ and oblateness for a given rotation rate, via their influence on stellar structure (top right). These determine the exterior space-time properties of the star, which measurably alter the properties of the radiation propagating from the NS surface, encoding in it information about the EoS and the associated microphysics.}	}
\vspace{-0.3cm}
\label{fig:eosphysics}
\end{figure}

Since we cannot directly sample the matter in the core of a NS, we must rely on indirect inference using sensitive observations of their exteriors.  Because the mass-radius ($M$-$R$) relation of NSs is strongly dependent on the EoS of the dense matter in their interior  \cite[see, e.g.,][]{Lattimer01,Lattimer05,Ozel09,Read09a,hebeler13}, measurements of the masses and radii of several NSs to a precision of a few percent using astrophysical observations can provide insight into the state of matter in their interior (Figure~\ref{fig:eosphysics}). 
This has motivated the development of a host of observational and data modeling techniques for inferring the masses and radii of NSs. A large subset of such methods rely on observations of the surface thermal radiation from neutron stars, because the properties of the observed photons are greatly effected by the immense gravity in the vicinity of the star \cite[e.g.,][and references therein]{ozel13,heinke13,miller13,potekhin14}, which in turn, is determined by the stellar $M$ and $R$. The X-ray portion of the electromagnetic spectrum ($\approx$0.1--30\,keV) is of particular interest in this regard, since it is where the bulk of the thermal radiation from NSs is observed.

\vspace{-0.5cm}
\section{X-ray Observational Techniques for EoS Constraints}
\vspace{-0.2cm}
{\bf Pulse profile modeling.} For a spinning NS that radiates X-rays from one or more hotter
regions on its surface (commonly referred to as ``hot spots''), its $R$ and $M$ can be estimated by fitting model pulse profiles to the observed X-ray pulsations, because the properties of the observed pulsed signal are affected by $R$ and $M$ through the general and special relativistic effects on rotationally-modulated emission from surface hot spots. Three varieties of NSs with surface hot spots suitable for such analyses are: rotation-powered millisecond pulsars, accretion-powered millisecond pulsars, and thermonuclear burst oscillation sources.  
The current \textit{Neutron Star Interior Composition Explorer} (\textit{NICER}; see
\cite{2016SPIE.9905E..1HG}) NASA X-ray timing mission is focusing on producing $R$ and
$M$ measurement of \textit{a few} radio millisecond pulsars that produce thermal radiation.
This pulse profile modeling technique (also known as waveform or light curve modeling) is mature, having been studied extensively over the last few decades \cite{Pechenick83,Miller98,Poutanen03,Poutanen06,Morsink07,Baubock13,Psaltis14,Algendy14,Nattila18}. Existing state-of-the-art models incorporate all practically important physics
(atmospheric radiation properties, gravitational light-bending, Doppler boosting, aberration, propagation time delays, and the effects of rotationally-induced stellar oblateness) with a very high degree of accuracy. Interfacing these models with Baysian parameter estimation codes enables inference of either exterior space-time parameters or EOS parameters directly from energy-resolved pulse profile data \cite{Lo13,MillerLamb15,Riley18,Raaijmakers18}.  Efforts for the {\em NICER} mission have led to the development of highly efficient codes optimized specifically for this purpose.

\begin{figure}[tb]
\centering
\includegraphics[width=0.94\textwidth]{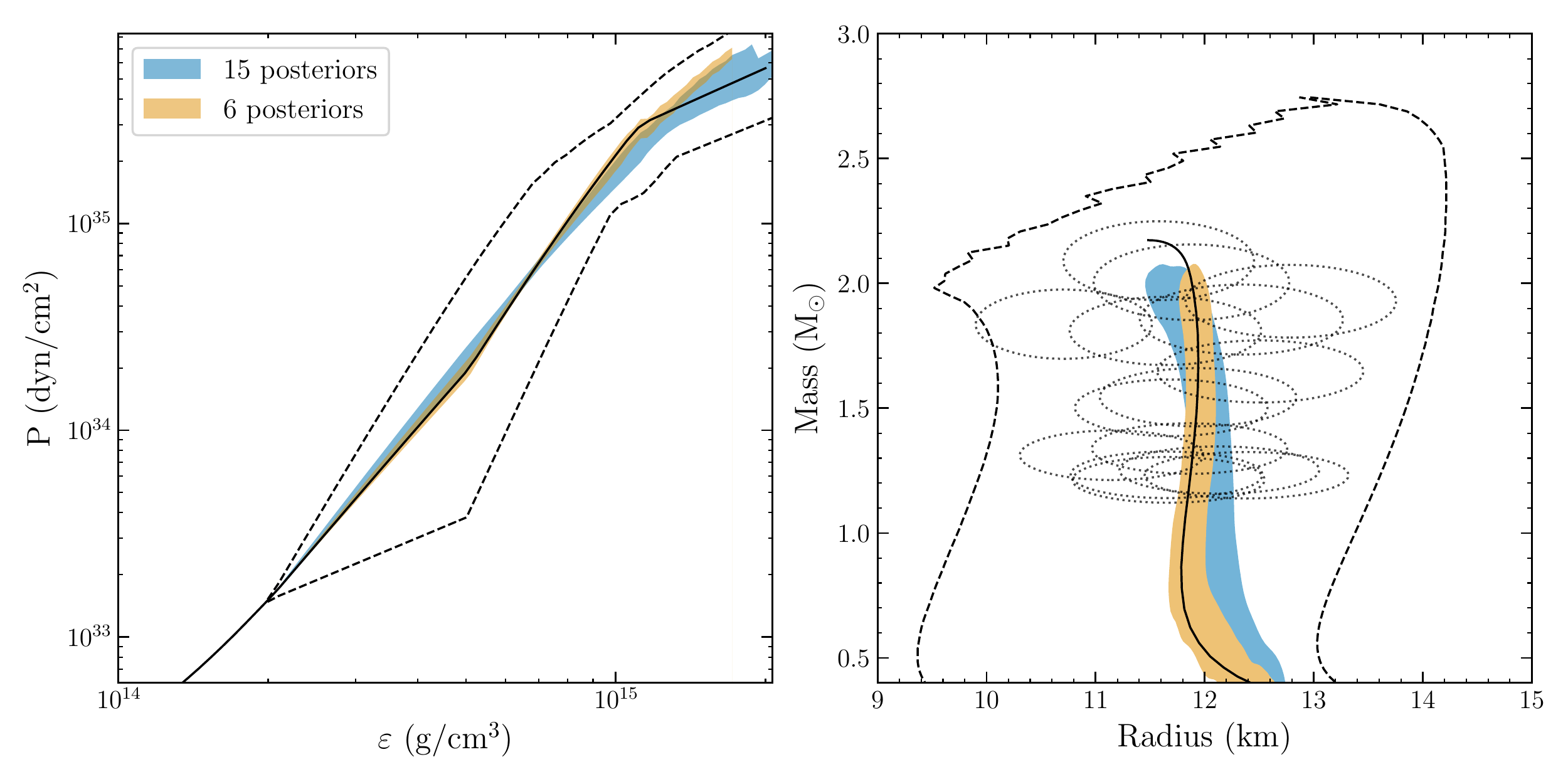}
\vspace{-0.2cm}
\caption{\small{Simulated constraints on the EoS and $M-R$ relation based on 15 expected measurements with \textit{STROBE-X}.  X-ray pulse profile modeling produces $M-R$ posteriors with $\pm 5$\% accuracy (dashed ellipses) scattered around an underlying $M-R$ relation (black line). 
The corresponding EoS is shown on the left hand panel, with the dashed lines indicating the range permitted by current models.  A piecewise-polytropic parameterization is assumed for the EoS with fixed transition densities.  Following the procedure outlined in \cite{Greif18}, the 1$\sigma$ constraints resulting from inference using these 15 measurements are shown by the blue band. The orange band shows the results for a choice of 6 of these stars, based on ensuring an even spread over the mass range $1.2 - 2.0$ M$_{\odot}$, for deeper observations that result in $\pm 2$\% accuracy posteriors (not shown)}. See \cite{2019arXiv190303035R} for further details.}	\label{fig:eosconstraints}
\end{figure}

{\bf Extremely rapid rotators.} X-ray searches for new transiently accreting pulsars have the potential to place  strong and clean (i.e., systematics-free) EoS constraints through the discovery of one or more NSs with spin rate $\gtrsim$1 kHz. Each theoretical EoS has an associated prediction for the maximum spin rate before break-up (e.g, \cite{Haensel09}), so such NSs could potentially rule out a swath of proposed EoS models. X-ray timing offers a particularly promising approach, since there is no evidence yet that the spin distribution of accreting millisecond pulsars drops off at high spin rates as seen in the radio pulsar sample \cite{Watts16}. Such rapidly spinning stars are likely to exhibit short-lived, intermittent X-ray pulsations, requiring hard X-ray instruments with fast timing capabilities and high instantaneous sensitivity.

{\bf Absorption line spectroscopy.} If photospheric absorption lines are detected from a NS, their gravitational redshift would provide a measure of the stellar compactness $M/R$ \cite{Burbidge_1963}.  A sufficiently precise measurement of the line shape would then also allow separate determination of $M$ and $R$ through the line broadening physics \cite{Paerels_1997, Loeb_2003, Chang_etal_2005}. Absorption lines in NSs can arise from either atomic transitions or magnetic effects \cite[see, e.g.][for a review]{ozel13}.  In the absence of an independent determination of magnetic field strength, magnetic absorption features are unsuitable for redshift measurements.  Atomic line features have yet to be detected from a NS surface; the most promising targets are thermonuclear (type-I) X-ray bursters, which generate bright episodic emission from the NS surface, supply heavy elements to the photosphere through accretion, and are uncontaminated by broad cyclotron lines owing to their weak magnetic field strengths.  Their one drawback is that most bursters have spin rates in the 200--600\,Hz~range, rendering photospheric lines undetectable by current instruments due to rotational broadening \cite{Ozel_Psaltis_2003}.  However, at least one slowly rotating (11 Hz) burster is now known \cite{Papitto_etal_2011}, and the greatly improved spectroscopic sensitivity of future missions like {\em Athena} and {\em Lynx} will extend access to at least some rapid rotators as well. 

{\bf Radius expansion bursts.} A subset of thermonuclear bursts are powerful enough to exceed the Eddington critical luminosity, lifting the photosphere above the NS surface \cite{Tawara_etal_1984, Lewin_etal_1984}. For sources whose distance is known (e.g., globular cluster sources), careful modeling of these so-called photospheric radius expansion (PRE) bursts permits measurement of both $M$ and $R$ \cite[see, e.g.,][]{Ozel_etal_2010}. However, because PRE events are accompanied by a drop in blackbody temperature, most of their luminosity lies outside the bandpass of most previous X-ray missions, leading to substantial uncertainties in the measured Eddington flux \cite[e.g.,][]{Kuulkers_etal_2003, Guver_etal_2012}. There are two different approaches. In one, the Eddington flux is measured at the moment of ``touchdown'', when the photosphere settles back onto the surface as the burst cools \cite{Ozel_etal_2009, Steiner_etal_2010}.  In the other, the spectral evolution of the burst's entire cooling tail is modeled \cite{Nattila17}. The energy band of the current {\em NICER} mission, which extends as low as 0.1~keV, is well suited to such measurements.  Future missions with both low-energy coverage and larger area will be able to measure the photospheric evolution in detail. Besides determining the touchdown point with great accuracy, the evolution itself will provide additional determinations of $M$ and $R$, as both the Eddington limit and the gravitational redshift at each radius depend upon these parameters \cite{Damen_etal_1990}. PRE bursts may also briefly uncover metals in burst ashes \cite{Weinberg_etal_2006, intZand_Weinberg_2010, YuWeinberg18}. These metals may produce gravitationally redshifted absorption features that can be used to measure $M$ and $R$, as described above.

{\bf Quiescent X-ray transients.} For a NS radiating uniformly from the entire surface, one can derive constraints on $M$ and $R$ from spectral fits to its X-ray emission if the temperature, composition of the outer atmosphere, and its distance are known and the magnetic field is sufficiently weak ($\ll 10^{10}$\,G)  so as not to affect the opacity or temperature distribution on the NS surface. These criteria are met in  quiescent low-mass X-ray binaries (qLMXBs) containing NSs: in particular, those located in globular clusters, to which the distances are well-determined \cite{Rutledge_etal_2002,Guillot_et_al_2013,Bogdanov_et_al_2016,Steiner_etal_2018}; plus additional targets in the field of the Galaxy with reliable parallaxes obtained with {\em GAIA}. Due to severe source crowding in the cores of globular clusters, effective studies of these targets require high-angular resolution soft X-ray imaging capabilities.

\vspace{-0.4cm}
\section{X-ray Observatories in the 2020s and Beyond}
\vspace{-0.2cm}
Observations at X-ray energies offer various means for obtaining strong constraints on the allowed dense matter equation of state. While current observatories have made important headway, they lack the required capabilities to fully exploit the information about the dense matter EoS encoded in the observed X-ray emission. Therefore, accomplishing this important undertaking requires a new generation of telescopes with at least an order-of-magnitude improvement in sensitivity across the soft and hard X-ray bands relative to existing observatories. 

The {\em Spectroscopic Time-Resolving Observatory for Broadband Energy X-rays} ({\em STROBE-X}) is a proposed probe-class mission specifically designed for X-ray timing and spectroscopy in the 0.2--30 keV band, with a 2--5\,m$^2$ collecting area, superb spectral and temporal resolution, and rapid slew capabilities \cite{2019arXiv190303035R}. {\em STROBE-X} would have the capabilities to employ multiple observational techniques to study the EoS \cite[see Table 1 and][for a review]{Watts16}.  The pulse profile modelling that can be done at present for a few pulsars with \nicer{} (sufficient to provide a proof of concept, \cite{Bogdanov13}) will be feasible for $\sim$20 pulsars with {\em STROBE-X} (Figure~\ref{fig:eosconstraints}), with more source classes becoming accessible through the hard X-ray band not covered by \textit{NICER}.  

The {\em Advanced X-ray Imaging Satellite} (\textit{AXIS}) is a proposed probe-class soft X-ray (0.1--10 keV) mission, with $\sim$$0.3''$ angular resolution imaging capabilities and a collecting area of $\approx4000$ cm$^{-2}$ around 1 keV \cite{2018SPIE10699E..29M}, with possible deployment in the late 2020s. These design features are well suited for spectroscopic studies of NS in crowded regions, such as qLMXBs in globular clusters.

 The {\em Advanced Telescope for High-ENergy Astrophysics} ({\em Athena}) is ESA's next-generation flagship X-ray observatory, slated for launch in the early 2030s  \cite{2013arXiv1306.2307N}. Its high sensitivity, high spectral resolution, and high count rate capability will enable studies of NSs where greatly improved spectroscopic sensitivity is desirable, in particular for PRE burst systems and qLMXBs \cite{2013arXiv1306.2334M}.
 
 The \textit{Lynx X-ray Observatory} \cite{2018arXiv180909642T} is a concept soft X-ray facility for possible selection as a NASA Large Strategic Science Mission with a target launch of $\sim$2035. It will have a novel combination of $\approx$2\,m$^2$ collecting area around 1 keV, high spectral resolution (through the use of a microcalorimeter and dispersion gratings), and focusing optics that enable sub-arcsecond imaging. With these performance characteristics, {\em Lynx} would be able to produce potentially strong NS EoS constraints through high fidelity spectra of both quiescent and bursting NS LMXBs, especially those in the dense confines of globular clusters and the Galactic center where such systems are overabundant, so high-angular resolution is essential.

\begin{table}[tbh]
    \centering
    \small
    \begin{tabular}{l|c|c}
\hline
\hline
    \textbf{Technique}  & \textbf{Source class} & \textbf{Observatories} \\
    \hline
    Pulse profile modeling  & rotation-powered MSPs & \textit{STROBE-X}, \textit{Athena} \\
                            & accretion-powered MSPs &        {\em STROBE-X}          \\
                            & bursting NS LMXBs  &        \textit{STROBE-X}     \\
                            \hline
    Extremely rapid rotation           &    accretion-powered MSPs    & \textit{STROBE-X}   \\
    \hline
    Radius expansion bursts  &    bursting NS LMXBs    & \textit{STROBE-X}   \\    
    \hline
    Absorption line spectroscopy     & bursting NS LMXBs & \textit{Athena}, \textit{Lynx} \\
    \hline
    Continuum spectroscopy     & quiescent NS LMXBs & \textit{AXIS}, \textit{Athena}, \textit{Lynx} \\
    \hline
    \end{tabular}
    \caption{\small{Summary of measurement techniques and NS source classes suitable for dense matter EoS constraints and the proposed X-ray observatories that have the capabilities to carry out these investigations.} }
    \label{tab:summary}
\end{table}

\vspace{-0.5cm}
\section{Conclusions}
\vspace{-0.2cm}
Table~\ref{tab:summary} summarizes the various techniques and NS source classes that can be employed to provide constraints on the dense matter EoS using X-ray observations, and the future planned observatories with the capabilities to accomplish the desired measurements. The ability to target multiple source classes is crucially important, since it provides verification of the measurement techniques, allowing characterization and mitigation of systematic errors.  For instance, a number of NSs exhibit both accretion-powered pulsations and thermonuclear burst oscillations, permitting pulse profile modelling for the same source using two different types of hot spot. Additionally, for the bursting sources, conducting spectroscopic modeling of the burst cooling tail can offer additional cross-checks of techniques \cite[see, e.g.][]{Nattila17}. By targeting more NSs, it will be possible to sample the EoS across a wider range of core densities.  This will map the EoS more fully, probing any potential phase transitions with finer resolution, and will move us out of the regime where EoS model parameter inference may be prior-dominated \citep[see for example][]{Greif18} . 

Identifying the heaviest NSs through radio pulsar timing provides additional potential for constraining the EoS \cite[see, e.g.,][]{Demorest_etal_2010}. 
Thus, maintaining good capabilities for finding pulsars and obtaining their timing solutions may contribute to our understanding of the NS EoS (see, e.g., the related White Paper by Fonseca et al.).
The Advanced LIGO and VIRGO gravitational wave observatories have now made the first direct detection of a binary NS merger \cite{Abbott18_BNS}.  With future runs, they may be able to constrain $R$ to $\sim$10\% using a few tens of detections \cite{Read13,Lackey15}.
Collectively, the resulting complementary measurements hold the promise to provide definitive empirical constraints on the true nature of the densest matter in the Universe. 
Pulsars and quarks were discovered within four years of the birth of X-ray astronomy, over half a century ago. The next generation of X-ray telescopes will bring these fields together, achieving breakthrough insights in nuclear physics using neutron stars.
 
\clearpage

\vspace{-0.4cm}
\bibliographystyle{aasjournal}
\bibliography{wp}

\end{document}